\documentclass[12pt]{article}

\usepackage{epsfig}
\def\nn{\nonumber}
\newcommand{\ovl}[1]{\overline{#1}}
\renewcommand{\d}{{\rm d}}
\def\adj{{\phantom{.}}}   
\def\als{\alpha_s}
\def\gev{\,{\rm GeV}}
\newcommand{\lsim}{\raisebox{-3pt}{$\,\stackrel{\textstyle <}{\sim}\,$}}

\begin{document}
\thispagestyle{empty}

\begin{flushright}
WU B 00-20 \\
hep-ph/yymmxxx\\
November 200\\[20mm]
\end{flushright}
\begin{center}
{\bfseries SKEWED PARTON DISTRIBUTIONS AND WIDE-ANGLE EXCLUSIVE SCATTERING}

\vskip 5mm

P.\ Kroll

\vskip 5mm
 
{\small {\it
Fachbereich Physik, Universit\"at Wuppertal, D-42097 Wuppertal, Germany
}}
\end{center}
\vskip 15mm
\begin{center}
\begin{minipage}{150mm}
\centerline{\bf Abstract}
The overlap representation of skewed parton distributions (SPDs) is discussed
and applications to wide-angle Compton scattering and electroproduction 
of mesons are presented. The amplitudes for these processes factorise 
into parton-level subprocess amplitudes and form factors representing 
$1/x$-moments of SPDs. 
\end{minipage}
\end{center}
\vskip 28mm
\begin{center}
Invited talk presented at the XV International Seminar on High Energy
Physics Problems, Dubna (September 2000)
\end{center}
\newpage

\thispagestyle{empty}
\setcounter{page}{1}
\begin{center}
{\bfseries SKEWED PARTON DISTRIBUTIONS AND WIDE-ANGLE EXCLUSIVE SCATTERING}

\vskip 5mm

P.\ Kroll

\vskip 5mm
 
{\small {\it
Fachbereich Physik, Universit\"at Wuppertal, D-42097 Wuppertal, Germany
}
}
\end{center}
\vskip 5mm
\begin{center}
\begin{minipage}{150mm}
\centerline{\bf Abstract}
The overlap representation of skewed parton distributions (SPDs) is discussed
and applications to wide-angle Compton scattering and electroproduction 
of mesons are presented. The amplitudes for these processes factorise 
into parton-level subprocess amplitudes and form factors representing 
$1/x$-moments of SPDs. 
\end{minipage}
\end{center}
\vskip 8mm
\section{Introduction}
Interactions of protons with high energy electromagnetic probes,
either electrons (i.e., virtual photons) or real photons, are desribed
by the handbag diagram shown in Fig.\ \ref{fig:handbag}. Depending on the
virtuality of the incoming photon, $Q^2$, and on the momentum transfer
from the incoming to the outgoing proton, $t$, different processes are
described by the handbag diagram: For forward scattering, $t=0$, with
highly virtual photons, $Q^2\gg \Lambda^2$, (where $\Lambda$ is a
typical hadronic scale of order 1 GeV) the cutted diagram
represents the total cross section for the absorption of virtual
photons by protons. This is the domain of deep inelastic lepton-proton
scattering from which process we learned about the ordinary
unpolarised, $q(x)$, and polarised, $\Delta q(x)$, parton distribution
functions (PDFs). Considering for definiteness outgoing real photons,    
the handbag diagram describes deeply virtual Compton scattering 
for large $Q^2$ and small $-t$ and, for large $-t$ (and $-u$) but
small $Q^2$, real and virtual wide-angle Compton scattering.   
Common to all these processes are soft proton matrix elements which
represent generalised PDFs or, as termed frequently, SPDs.
The same matrix elements also occur in the form
factors of the proton and in deeply virtual as well as wide-angle 
electroproduction of mesons. In the following I am going to discuss
properties of the SPDs and their representation in terms of light-cone 
wave function (LCWF) overlaps. The role of the SPDs in wide-angle
exclusive scattering will also be explained.  
\section{Skewed parton distributions}
\begin{figure}[t]
\parbox{\textwidth}{\begin{center}
   \psfig{file=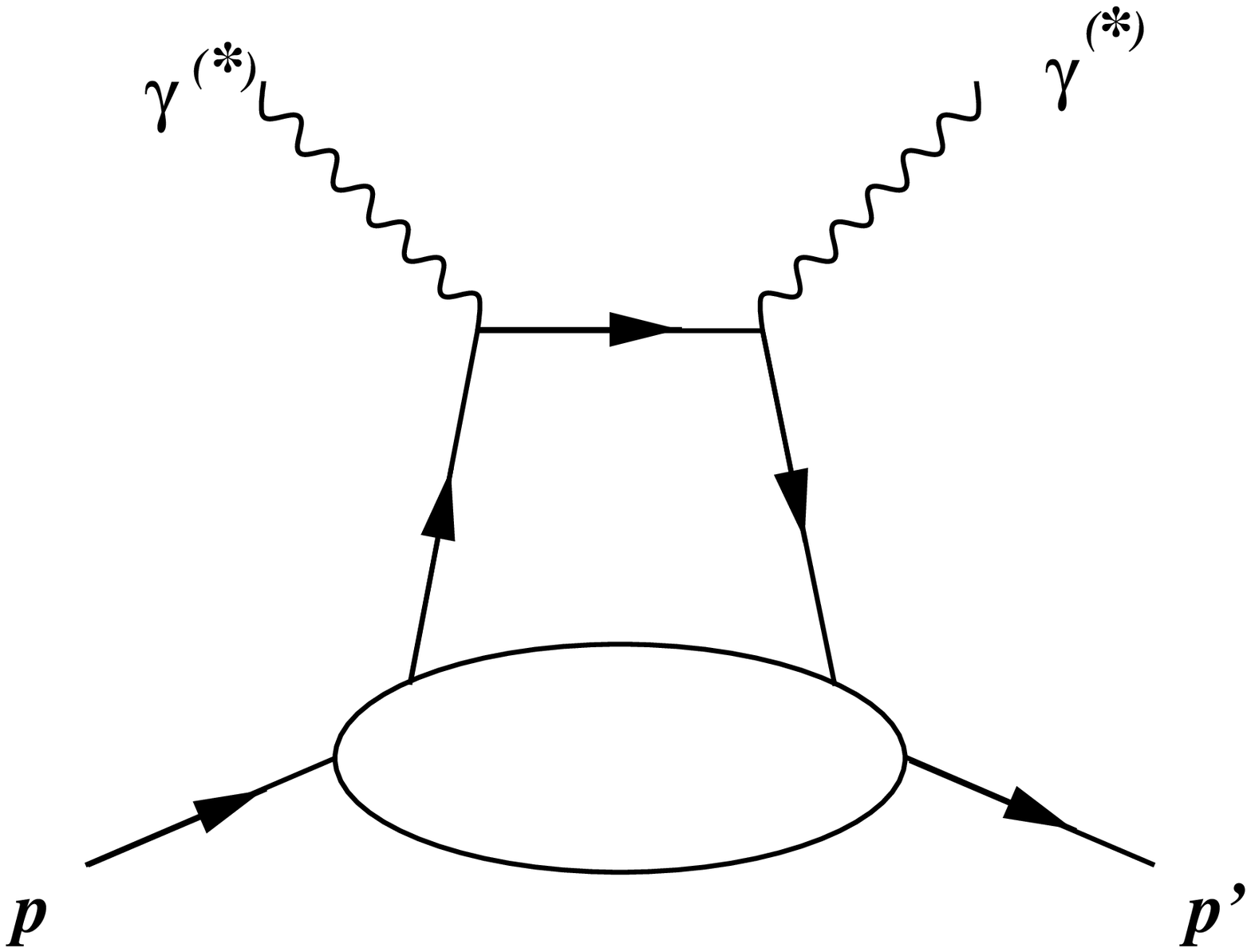,%
          bbllx=45pt,bblly=230pt,bburx=545pt,bbury=610pt,%
           width=4.0cm,clip=} \hspace{2cm}
   \psfig{file=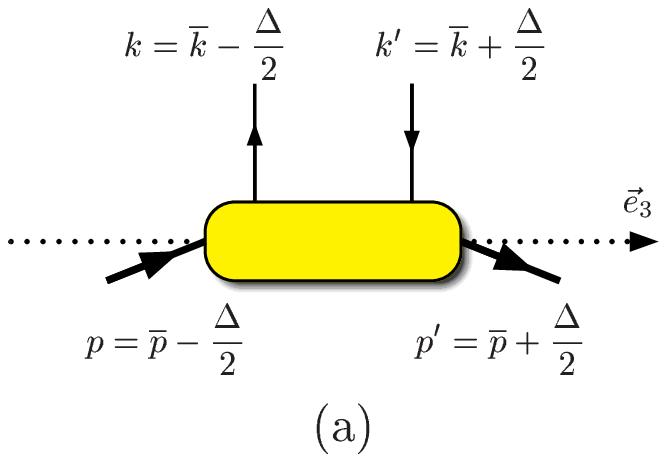,%
          bbllx=190pt,bblly=460pt,bburx=380pt,bbury=580pt,%
           width=4.5cm,clip=}
\vspace{-0.5cm}
\end{center}}
\caption{The handbag diagram (left) and the kinematics for SPDs (right).}
\label{fig:handbag}
\end{figure}

Parameterizing the kinematics of the SPD as in Fig.\ \ref{fig:handbag}, 
one can define the following fractions of light-cone plus
momentum components
\begin{equation}
\xi=\frac{(p-p^\prime)^+}{(p+p^\prime)^+}\,, \qquad 
\bar{x}=\frac{(k+k^\prime)^+}{(p+p^\prime)^+}\,, \qquad
x=\frac{\bar{x}+\xi}{1+\xi}\,, \qquad
x^\prime=\frac{\bar{x}-\xi}{1-\xi}\,,
\label{eq:fractions}
\end{equation}
where $\xi$ is termed the skewedness parameter. $x (x^\prime)$ is the
momentum fraction the emitted (absorbed) parton carries. The SPDs
$H^q(\bar{x},\xi;t)$ and $E^q(\bar{x},\xi;t)$ for a quark of flavour
$q$ are defined as the Fourier transform of a bilocal product of field
operators sandwiched between proton states \cite{mue:94}
  
\begin{eqnarray} 
\label{eq:quarkSPDdef}
{\cal H}_{\lambda'\lambda}^q &\equiv&
\frac{1}{2\sqrt{1-\xi^2}} \;
\sum_c
\int\frac{\d z^-}{2\pi}\;e^{i\,\bar x\,\bar p^{\,+}z^-}\;
\langle p^{\,\prime},\lambda'|
   \,\overline\psi_q^{\,c}(-\bar z/2)\,\gamma^+\,\psi_q^{\,c}(\bar z/2)\,
                             |p,\lambda\rangle
\nn\\[1\baselineskip]
&=&
 \frac{\overline u(p^{\,\prime},\lambda')\gamma^+ u(p,\lambda)}
      {2\bar p^{\,+}\sqrt{1-\xi^2}}\; 
H^q(\bar x,\xi;t)
+\frac{\ovl u(p^{\,\prime},\lambda')
         {i\sigma^{+\alpha}\Delta_\alpha} u(p,\lambda)}
      {4m\,\bar p^{\,+}\sqrt{1-\xi^2}}\; 
E^q(\bar x,\xi;t)\,,
\end{eqnarray} 
(in $A^+=0$ gauge, $t=\Delta^2$). $\lambda$, $\lambda^\prime$ denote
the proton helicities and $\bar{z}$ is a shorthand notation for
$[0,z^-,{\bf 0}_\perp]$. The decomposition (\ref{eq:quarkSPDdef})
defines the two SPDs in analogy to the Dirac, $F_1$, and Pauli, $F_2$,
form factors of the proton. The corresponding matrix element of
$\gamma^+ \gamma_5$ defines two further SPDs, $\widetilde{H}^q$ and
$\widetilde{E}^q$. For $\xi < \bar{x} \leq 1$ the SPDs describe the
emission of a quark with momentum fraction $x > 0$ from the proton
and the absorption of a quark with momentum fraction $x^\prime >
0$. In the region $-\xi \le \bar{x} \le \xi$ the proton emits a
quark-antiquark pair ($x \geq 0$, $x^\prime \leq 0$) and is left as a
proton. Finally, for $-1 \leq \bar{x} < -\xi$ the SPD describes the
emission and absorption of antiquarks ($x, x^\prime <
0$). Antiquark SPDs can be defined by $H^{\bar{q}}(\bar{x}, \xi; t) =
- H^q(-\bar{x}, \xi; t)$ and so on. The extension to gluon SPDs is
straightforward.

Reduction formulas relate the SPDs to the ordinary PDFs:
\begin{eqnarray}
\label{eq:red1}
H^q(\bar{x},0;0) &=& q(\bar{x}), \phantom{x} \qquad  
\widetilde{H}^q(\bar{x},0;0)\; = \;\Delta q(\bar{x}) \nn \\
H^g(\bar{x},0;0) &=& \bar{x} g(\bar{x}), \qquad 
\widetilde{H}^g(\bar{x},0;0)\; = \;\bar{x} \Delta g(\bar{x})\,.
\end{eqnarray}
The other SPDs, $E^q$ and $\widetilde{E}^q$, are not accessible to
deep inelastic lepton-nucleon scattering.
Integrating the SPDs over $\bar{x}$ one obtains the
contribution of flavour $q$ quarks to the proton form factors,
e.g.,
\begin{equation}
\label{eq:red2}
 F_1^q(t) = \langle \bar{x}^0 \rangle = \int_{-1}^1 \d\bar{x}\, 
                                 H^q(\bar{x},\xi;t) \,.
\end{equation}
The $\langle \bar{x}^0 \rangle$-moments of the other
SPDs, $E^q$, $\widetilde{H}^q$ and $\widetilde{E}^q$, provide
contributions to the Pauli, the axial, and the
pseudoscalar form factors, respectively. Multiplying, for
instance, $F^q_1$ by the appropriate electric charges and summing over
all flavours, one obtains the full Dirac form factor.

As shown in \cite{DFJK1,DFJK3} (see also \cite{bro:00}), the SPDs
possess a representation in terms of LCWF overlaps which is a
generalization of the famous Drell-Yan formula for the electromagnetic
form factors \cite{DY}. Its derivation starts from a Fock
decomposition of the proton state    
\begin{equation}
\label{eq:Fockstate}
\left|p,\lambda\right\rangle = \sum_{N,\beta} 
\int [{\d} x]_N [{\d}^2 {\bf k}_\perp]_N\;
\Psi_{N,\beta}^\lambda(r) \;
\left|N,\beta;k_1,\ldots,k_N \right\rangle \,,
\end{equation} 
where $\beta$ labels the different $N$-particle Fock states and
$[{\d}x]_N[{\d}^2 {\bf k}_\perp]_N$ is the $N$-particle integration
measure. In the framework of light-cone quantisation one can show that,
in the region $\xi < \bar{x} \leq 1$, 
the SPDs are represented by the diagonal $N\to N$ overlaps
\begin{equation} 
{\cal H}_{\lambda'\lambda}^{q(N\to N)} =
\sqrt{1-\xi^2}^{\,1-N}\;
\sum_{\beta=\beta'}\;
\sum_{j} \delta_{s_j q}\;
\int [\d\bar x]_N [\d^2 \bar{\bf k}_\perp]_N \;
\delta\left(\bar x-\bar x_j\right)\;
\Psi_{N,\beta'}^{*\,\lambda'}(\hat r')\,
\Psi_{N,\beta}^\lambda(\tilde r)\,,
\label{eq:quarkSPD}
\end{equation} 
where the arguments of the LCWFs read ($j$ labels the active quark,
$i\neq j$ the spectators)
\begin{eqnarray}
\label{eq:spect-args}
\tilde x_i= \frac{\bar x_i}{1+\xi} \,, \quad 
\tilde{\bf k}_{\perp i}=\bar{\bf k}_{\perp i}
                        +\frac{\bar x_i}{1+\xi}\,
                         \frac{{\bf\Delta}_\perp}{2}\,, \quad 
\tilde x_j=\frac{\bar x_j+\xi}{1+\xi} \,, \quad
\tilde{\bf k}_{\perp j}=\bar{\bf k}_{\perp j}
                        -\frac{1-\bar x_j}{1+\xi}\,
                         \frac{{\bf\Delta}_\perp}{2}\,, \nn\\[2mm]
\hat x'_i= \frac{\bar x_i}{1-\xi} \,, \quad 
\hat{\bf k}'_{\perp i}=\bar{\bf k}^\adj_{\perp i}
                        -\frac{\bar x_i}{1-\xi}\,
                         \frac{{\bf\Delta}_\perp}{2}\,, \quad
\hat x'_j=\frac{\bar x_j-\xi}{1-\xi} \,, \quad 
\hat{\bf k}'_{\perp j}=\bar{\bf k}^\adj_{\perp j}
                        +\frac{1-\bar x_j}{1-\xi}\,
                         \frac{{\bf\Delta}_\perp}{2}\,.
\end{eqnarray} 
The overlap representation of the matrix element $\widetilde{{\cal
H}}^{q(N\to N)}_{\lambda^\prime \lambda}$ is, except of an additional
factor $\rm{sign}(\mu_j)$ (where $\mu_j$ is the helicity of the active
parton), also given by (\ref{eq:quarkSPD}). For the gluon SPDs there is
an additional factor $\sqrt{\bar{x}^2 - \xi^2}$. The overlap
representation in the region $-1\leq\bar{x} < -\xi$ is given by 
(\ref{eq:quarkSPD}) too with the replacement of $\delta(\bar{x}-\bar{x}_j)$ 
by $\delta(\bar{x}+\bar{x}_j)$. The region $-\xi \le \bar{x} \le \xi$ 
is described by non-diagonal $N+1 \to N-1$ overlaps. I refrain from 
displaying that formula here and refer to Ref.\ \cite{DFJK3,bro:00}. 

In the region $\xi < \bar{x}\leq 1$ the overlap representation
satisfies the positivity constraints 
\begin{equation}
\left| {\cal H}^{q}_{\lambda^\prime \lambda} \right|
     \leq \frac{1}{\sqrt{1-\xi^2}}\;
          \sqrt{q(x)\;q(x^\prime)} \,,\qquad
\left| {\cal H}^{g}_{\lambda^\prime \lambda} \right|
     \leq \sqrt{\frac{\bar{x}^2-\xi^2}{1-\xi^2}}\;
          \sqrt{g(x)\;g(x^\prime)}\,, 
\label{eq:qbound}
\end{equation}
first derived in \cite{pir:99}. I.e.\ we obtain bounds for the combination 
$ H^{q(g)} - \xi^2/(1-\xi^2)\, E^{q(g)} = {\cal H}_{\lambda\lambda}$
and for $E^{q(g)}$ ($\propto {\cal H}_{\lambda -\lambda}$)
\cite{DFJK3}. In previous work \cite{pir:99,DFJK1} the
$E^{q(g)}$ term in the bound for proton helicity non-flip transitions
have been overlooked. The combinations ${\cal
H}^{q(g)}_{\lambda^\prime \lambda} \pm 
        \widetilde{{\cal H}}^{q(g)}_{\lambda^\prime \lambda}$ are
analogously bounded by the PDFs for definite helicities \cite{DFJK3}.

Considering the special case $\Delta_\perp=\xi=0$,
$\lambda^\prime=\lambda$, Eq.\ (\ref{eq:quarkSPD}) and the
corresponding one for $\widetilde{{\cal H}}^q_{\lambda\lambda}$ reduce
to the LCWF representation of the ordinary PDFs, $q$ and $\Delta
q$, respectively. Thus, the reduction formulas (\ref{eq:red1}) are 
automatically satisfied. Taking $\xi=0$ and integrating 
(\ref{eq:quarkSPD}) over $\bar{x}$ the ordinary Drell-Yan formula 
\cite{DY} is recovered. In the same manner LCWF representations of the 
other proton from factors are obtained.
\section{The soft physics approach}
\begin{figure}[t]
\parbox{\textwidth}{\begin{center}
   \psfig{file=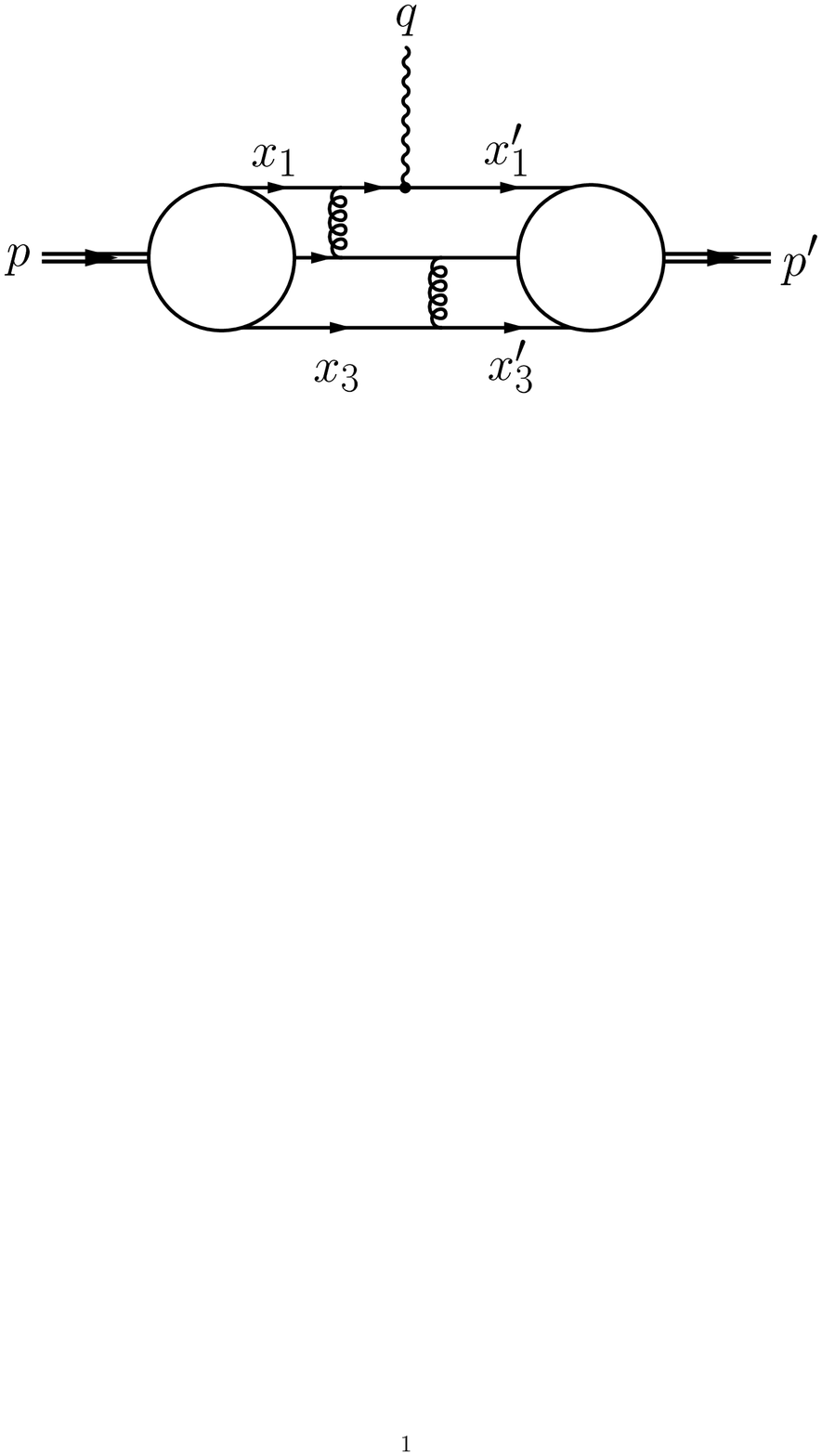,%
          bbllx=90pt,bblly=580pt,bburx=510pt,bbury=790pt,%
           width=5.0cm,clip=}\hspace{1.5cm}
   \psfig{file=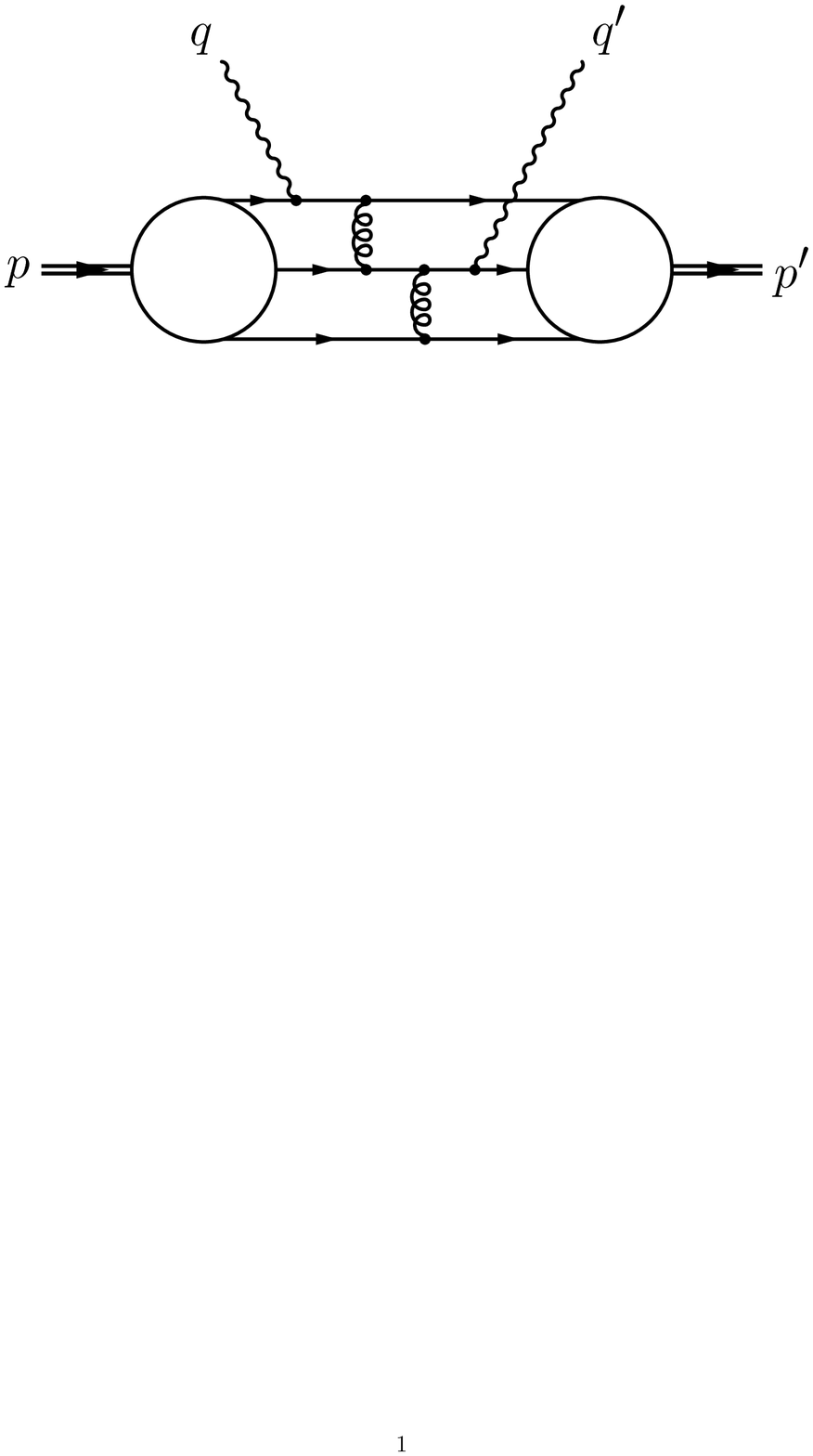,%
          bbllx=90pt,bblly=585pt,bburx=510pt,bbury=790pt,%
           width=5.0cm,clip=}
\vspace*{-0.5cm}
\end{center}}
\caption{Sample Feyman graphs for form factors (left) and Compton
scattering (right)  within pQCD.}
\label{fig:pqcd}
\end{figure}

\begin{figure}[t]
\parbox{\textwidth}{\begin{center}
   \psfig{file=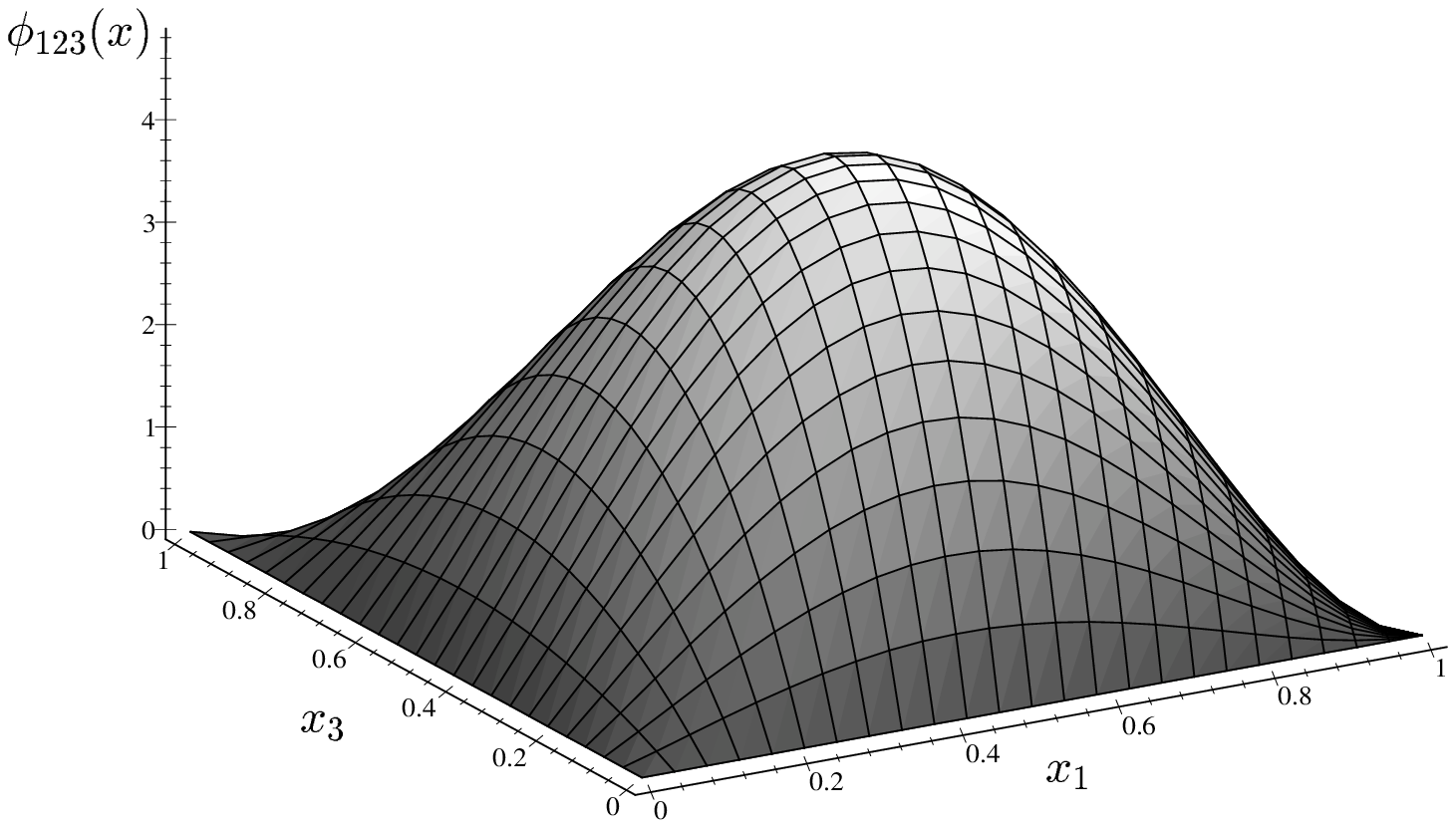,%
          bbllx=105pt,bblly=460pt,bburx=530pt,bbury=700pt,%
           width=5.0cm,clip=}\hspace{1.5cm}
   \psfig{file=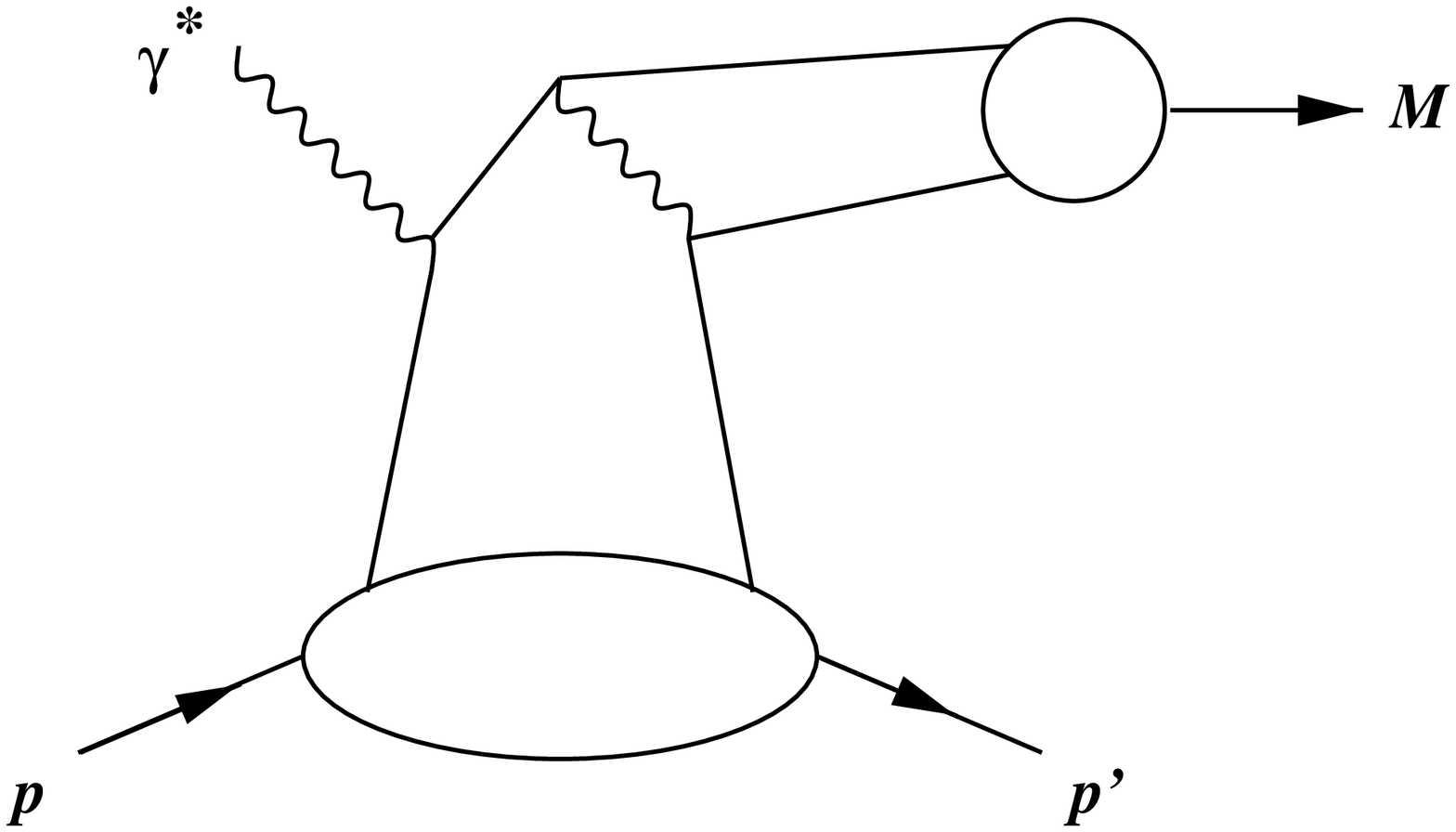,%
          bbllx=15pt,bblly=265pt,bburx=550pt,bbury=580pt,%
           width=4.5cm,clip=}
\vspace*{-0.5cm}
\end{center}}
\caption{The DA (\ref{eq:BKda}) taken from Ref.\ \cite{bol96} (left) and
a typical Feynman diagram for hard electroproduction of mesons (right).}
\label{fig:electro}
\end{figure}

Let us now turn to wide-angle exclusive scattering, defined by
$s,-t,-u \gg \Lambda$, $Q^2$ fixed. As is
well-known, for asymptotically large momentum transfer these processes
are controlled by the perturbative contributions where all partons the
proton is made off, participate in the hard scattering \cite{bro80},
see Fig.\ \ref{fig:pqcd}. The leading contributions are
generated by the valence Fock state with the quarks being
connected by the exchange of a minimal number of gluons. Higher Fock
states and the exchange of additional gluons provide power and  
$\als$ corrections to the leading contributions. 
It turned out however that the leading-twist perturbative
contributions to the proton form factor \cite{bol96,ber95} 
and to Compton scattering \cite{brooks} are ways below experiment unless
strongly asymmetric, i.e., end-point concentrated distribution
amplitudes (DAs) are used. (A DA represents a
LCWF integrated over tranverse momenta up to the factorisation
scale.) The use of asymmetric DAs is however inconsistent since the bulk of
a perturbative contribution evaluated from such a DA is accumulated in
the end-point regions where the assumptions of perturbative calculations break
down. It is also easy to see that a strongly asymmetric DA if combined with a
Gaussian transverse momentum dependence in a LCWF
\begin{equation}
\Psi^\lambda_{N\beta} \propto \exp{\left[-a_N^2 \sum^N_{i=1}
                                  k^2_{\perp i}/x_i\right]}\,,   
\label{eq:gaussian}
\end{equation}
leads to a large overlap contribution (evaluated by means of
(\ref{eq:quarkSPD}) and (\ref{eq:red2})) to the proton form factor
exceeding the data by a factor of about 5 to 7 \cite{bol96,isg}.
Moreover, the ordinary parton distributions, evaluated from such LCWFs
via (\ref{eq:quarkSPD}) and (\ref{eq:red1}) \cite{bol96,sch89},
are in sharp disagreement with the results obtained from analyses of
deep inelastic lepton-nucleon scattering \cite{GRV}. A consistent
description of form factors, parton distributions and, as we will see
below, Compton scattering \cite{DFJK1,rad98a,DFJK2} which
requires the inclusion of both the soft and the perturbative 
contributions, can only be achieved if DAs are used that are close to 
the asymptotic form $\Phi_{AS}=120 x_1 x_2 x_3$. An example is given 
in \cite{bol96} (see Fig.\ \ref{fig:electro})
\begin{equation}
\Phi_{123}^{BK} = \Phi_{AS}\, \frac12 (1+3 x_1)
\label{eq:BKda}
\end{equation}
which refers to the $u_+ u_- d_+$ configuration of the proton's
valence quarks. A recent instanton model study
\cite{dia00} seems to support the phenomenological DA
(\ref{eq:BKda}). As remarked in \cite{braun00} previous QCD sum rule 
studies \cite{COZ} if restricted to the first order Appell
polynomials, also provides a DA similar to the asymptotic one. The 
situation here seems to be quite analogous to the case of the pion 
where the analysis \cite{raulfs} of the CLEO data \cite{CLEO} on
the $\pi\gamma$ transition form factor as well as the E791 measurement
\cite{E791} clearly favour a form of the pion distribution amplitude
that is close to the asymptotic one. 

The perturbative contributions to the form factor and to Compton
scattering evaluated from the DA (\ref{eq:BKda}) amounts to less than
$10\%$ of experiment \cite{bol96,ber95,brooks}. The onset of the
perturbative regime is expected to lie above $t\simeq -100
\gev^2$. For wide-angle electroproduction of mesons there is no
reliable perturbative result available as yet. On account of the
experience with other exclusive reactions one may expect a small
perturbative contribution here too. Only the decays $J/\Psi, \Psi' \to
p\bar{p}$ are dominated by perturbative physics \cite{bol96,bol:98}.
   
The crucial question now arises - how to calculate the soft
contributions to wide-angle Compton scattering and electroproduction of
mesons? The relevant handbag diagram for Compton scattering is shown
in Fig.\ \ref{fig:handbag} and its generalization to electroproduction
in Fig.\ \ref{fig:electro} where a leading twist generation of the
meson is assumed. Note that for the production of flavour neutral
mesons active gluons have to be considered as well \cite{hanwen00}.
In Refs.\ \cite{DFJK1,hanwen00} the soft contributions to these
processes are defined through the assumption that the (soft) LCWFs are
dominated by parton virtualities $k_i^2, k'{}^2_i \lsim \Lambda^2$
and by intrinsic transverse momenta that satisfy $\tilde{k}^2_{\perp
i}/\tilde{x}_i,\: \hat{k}'{}^2_{\perp i}/\hat{x}'_i \lsim
\Lambda^2$. One can then show that the active partons are
approximately on-shell, collinear with their parent hadrons and they are carrying
momentum fractions of about unity. Thus, the physical situation is
that of a hard photon-parton scattering and a soft emission and
re-absorption of partons by the protons. The amplitudes therefore factorise
into subprocess amplitudes (either $\gamma^* q\to \gamma q$ or
$\gamma^* q \to M q$) and $1/\bar{x}$-moments of $\xi=0$ SPDs. The
helicity  amplitudes for Compton scattering, for instance, read
\cite{DFJK1}
\begin{equation}
{\cal M}_{\mu'+,\,\mu +}(s,t) \,=\, \;2\pi\alpha_{\rm em} \left[{\cal
    H}_{\mu'+,\,\mu+}\,(R_V(t) + R_A(t)) \,
    + \, {\cal H}_{\mu'-,\,\mu-}\,(R_V(t) - R_A(t)) \right ]\,.
\label{final}
\end{equation} 
Proton helicity flip is neglected. The subprocess amplitudes ${\cal
H}_{\mu'\nu,\mu\nu}$ are calculated for massless quarks to lowest order
perturbation theory \cite{DFJK1,hanwen00}. The soft form factors,
$R_V$ and $R_A$, for an active quark of flavour $q$, are defined by
\begin{equation}
R_V^{\,q}(t) = \int_{-1}^1
\frac{\d\bar x}{\bar x}\;H^q(\bar x, 0;t)\,, \qquad
R_A^{\,q}(t) = \int_{-1}^1 
\frac{\d\bar x}{\bar x}\;
\mbox{sign}(\bar x)\,
\widetilde{H}^q(\bar x, 0;t)\,.
\label{eq:cmoments}
\end{equation}
The full form factors are specific to the process under
consideration. All charged partons contribute to Compton scattering,
(e.g.\ $R_V(t) = \sum e_q^2 R_V^q(t)$)
while in electroproduction the meson selects its valence quarks from
the proton.
\section{Results}
\begin{figure} 
\parbox{\textwidth}{\begin{center}
       \psfig{file=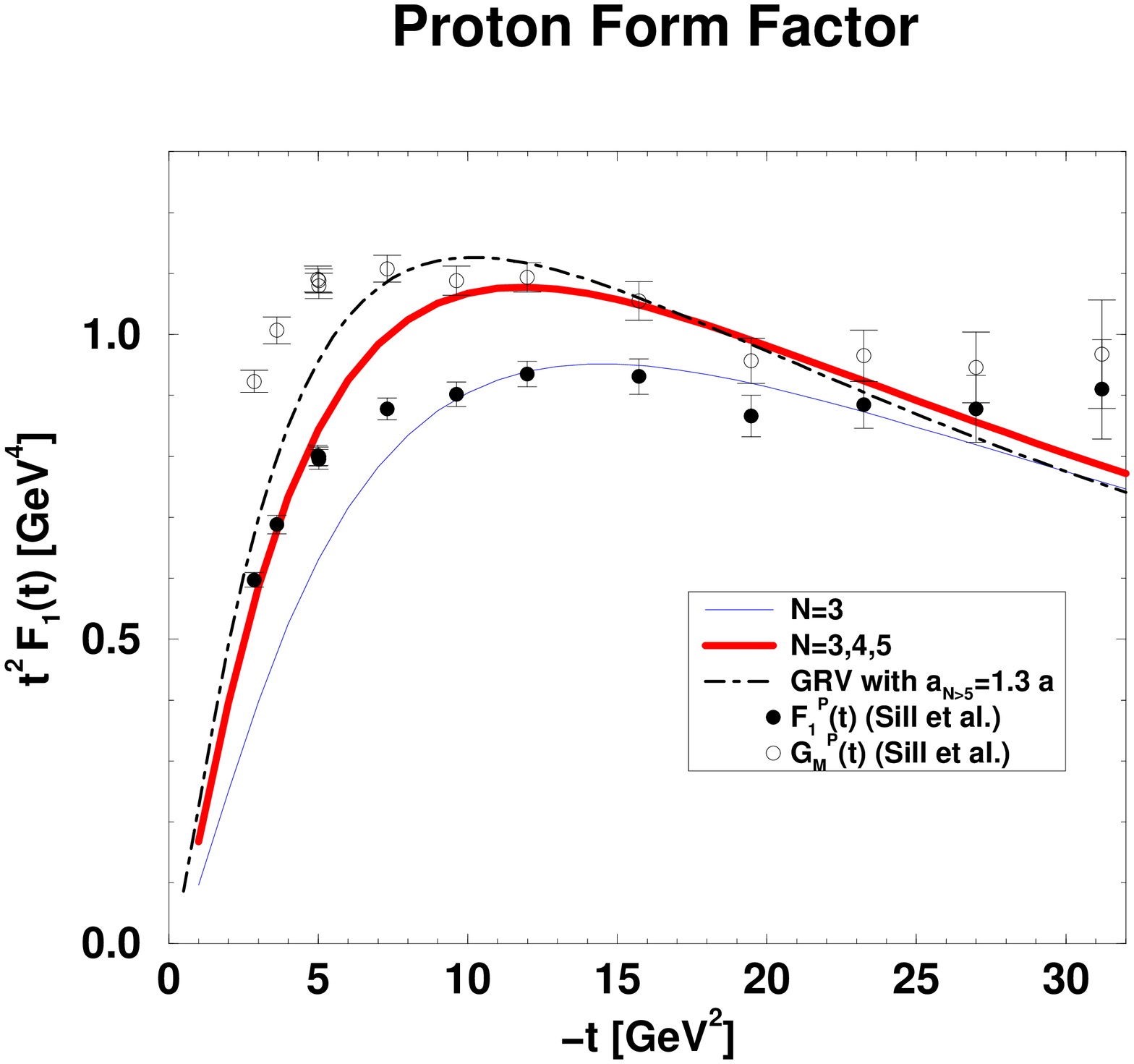, 
          bbllx=100pt,bblly=0pt,bburx=590pt,bbury=635pt,%
           width=5.5cm, angle=-90, clip=} \hspace{0.5cm}
       \psfig{file=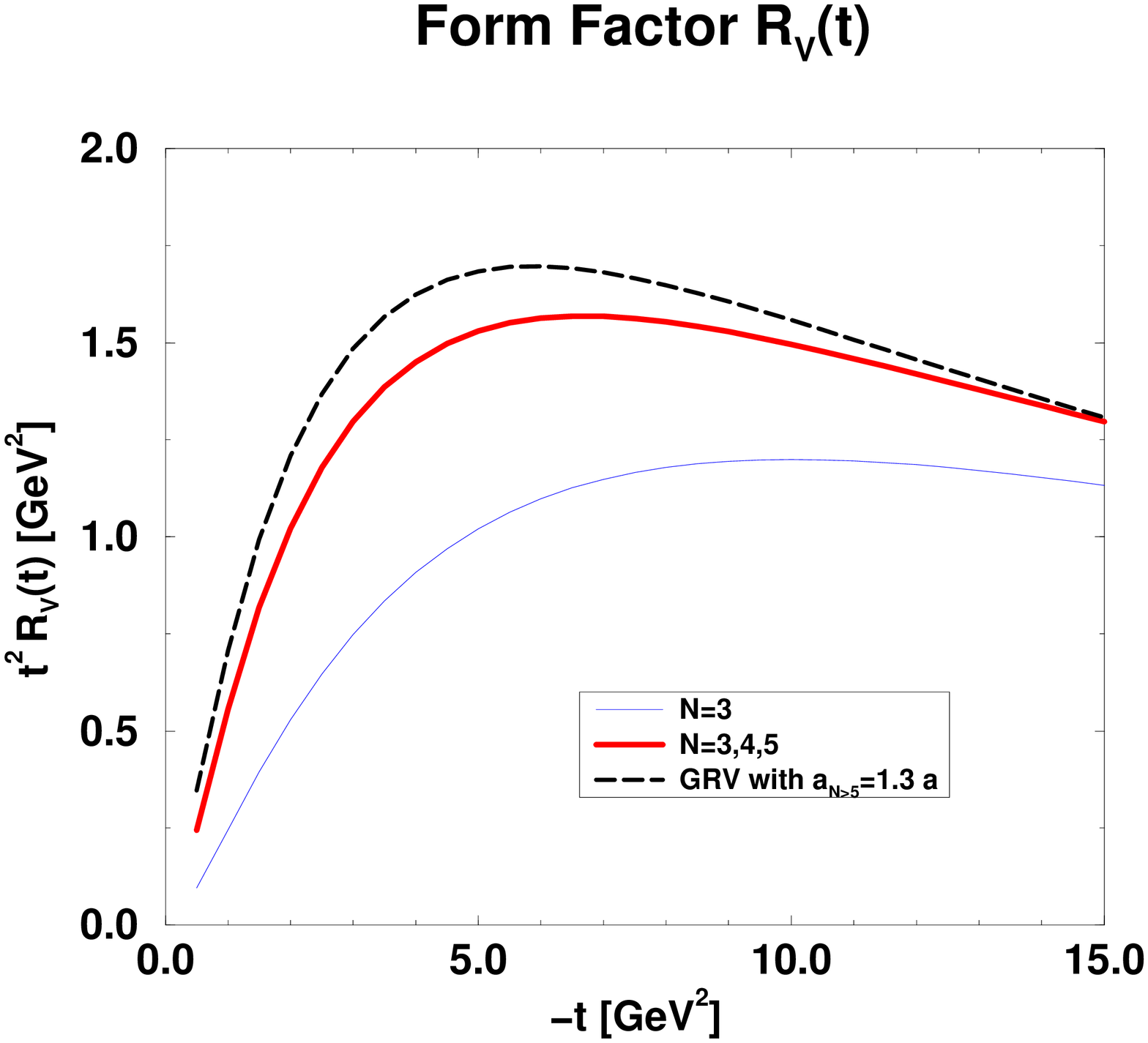, 
          bbllx=90pt,bblly=0pt,bburx=590pt,bbury=655pt,%
           width=5.5cm, angle=-90, clip=} 
\vspace*{-0.5cm}
\end{center}}
\caption{The Dirac (left) and the vector Compton (right) form factors
   of the proton as predicted by the soft physics approach
   \cite{DFJK1,bol96}. Data are taken from \cite{sil93}. The data on
   the magnetic form factor, $G_M$, are shown in order to demonstrate
   the size of spin-flip effects.} 
\label{fig:soft}
\end{figure}

In order to predict cross sections or polarization observables a model
for the new form factors, $R_{V}$ and $R_A$, is required. Parameterizing the
transverse momentum dependence of the LCWFs as in (\ref{eq:gaussian}),
which is in line with the central assumption of the soft physics
approach of restricted $k^2_{\perp i}/x_i$, and using a common
transverse size parameter ($a=a_N$) for simplicity, one can evaluate the
$\xi=0$ SPDs from (\ref{eq:quarkSPD}) and, without need for specifying
the $x$-dependences of the LCWFs, relate the results to the ordinary
parton distributions \cite{DFJK1}
\begin{equation}
H^q(\bar{x}, 0;t)=\, \exp{\left[\,\frac12\, a^2 t\,
               \frac{1-\bar{x}}{\bar{x}}\right]}\,\; q(\bar{x})\,,
\qquad
\widetilde{H}^q(\bar{x}, 0;t)=\, \exp{\left[\,\frac12\, a^2 t\,
               \frac{1-\bar{x}}{\bar{x}}\right]}\, \Delta q(\bar{x})
\,.
\label{xi0}
\end{equation}
Taking the parton distributions from one of the current analyses of
deep inelastic lepton-nucleon scattering, e.g., \cite{GRV}, using a 
value of 1 GeV$^{-1}$ for the transverse size parameter and evaluating
the moments of the SPDs according to (\ref{eq:red2}) and (\ref{eq:cmoments}), 
one finds reasonable results in fair agreement with experiment
\cite{DFJK1,rad98a,DFJK2}. Improvements are obtained by treating the
lowest three Fock states explicitly with specified $x$-dependences, the  
DA (\ref{eq:BKda}) for the valence Fock state and similar ones for
the next higher Fock states \cite{DFJK1,DFJK2}. The form factors $F_1$
and $R_V$, scaled by $t^2$, are displayed in Fig.\ \ref{fig:soft}.
The scaled form factors exhibit broad maxima and, hence, mimic the
dimensional counting rule behaviour in the $t$-range from about 5 to
15 GeV{}$^2$. This $t$-range is set by the transverse proton size. 
For very large momentum transfer the form factors turn gradually into
the soft physics asymptotics $\sim 1/t^4$. This is the region where the
perturbative contribution ($\sim 1/t^2$) takes the lead. 

\begin{figure}
\parbox{\textwidth}{\begin{center}
   \psfig{file=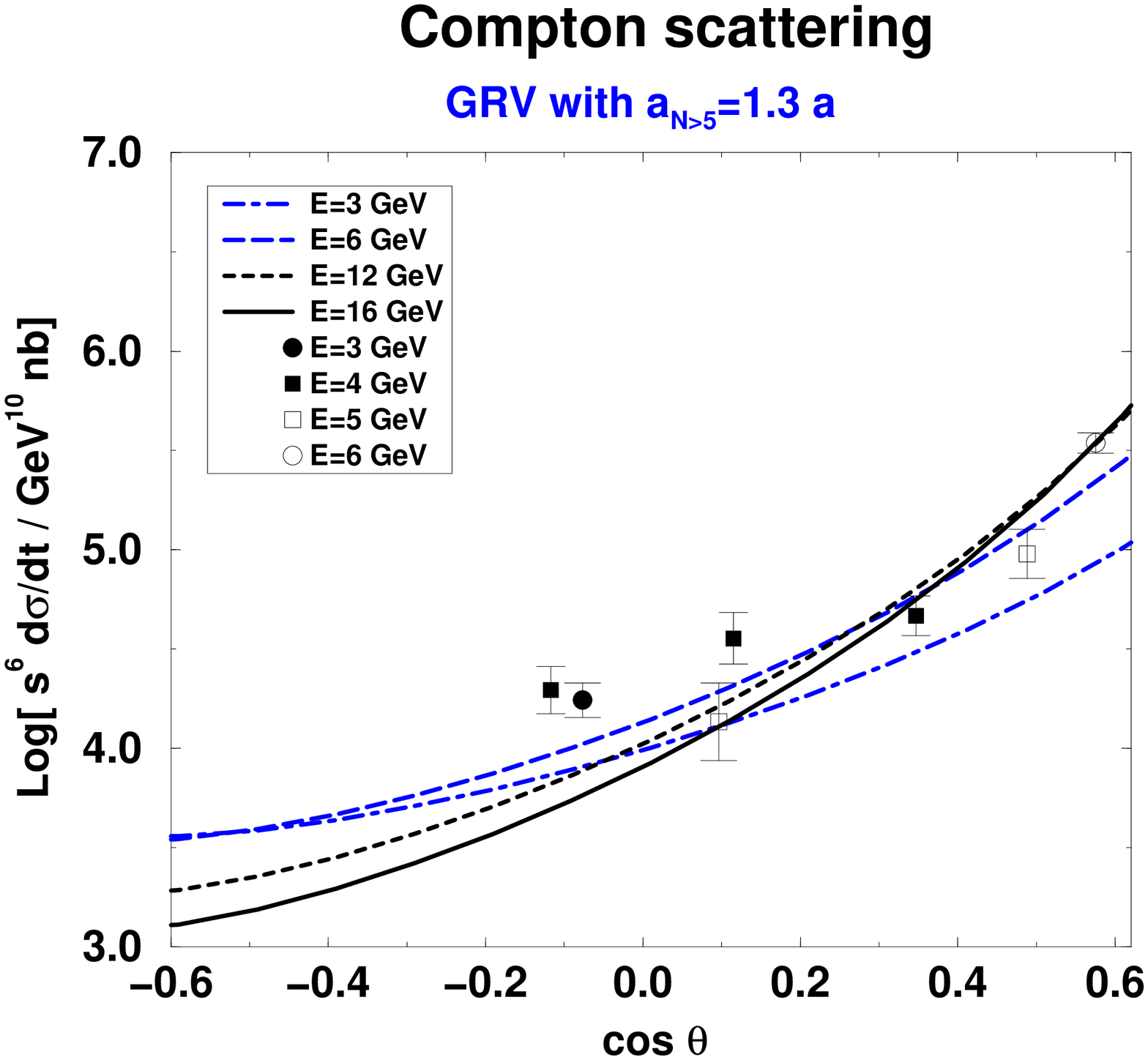, 
          bbllx=92pt,bblly=0pt,bburx=590pt,bbury=640pt,%
           width=5.7cm, angle=-90, clip=} \hspace{1cm}
   \psfig{file=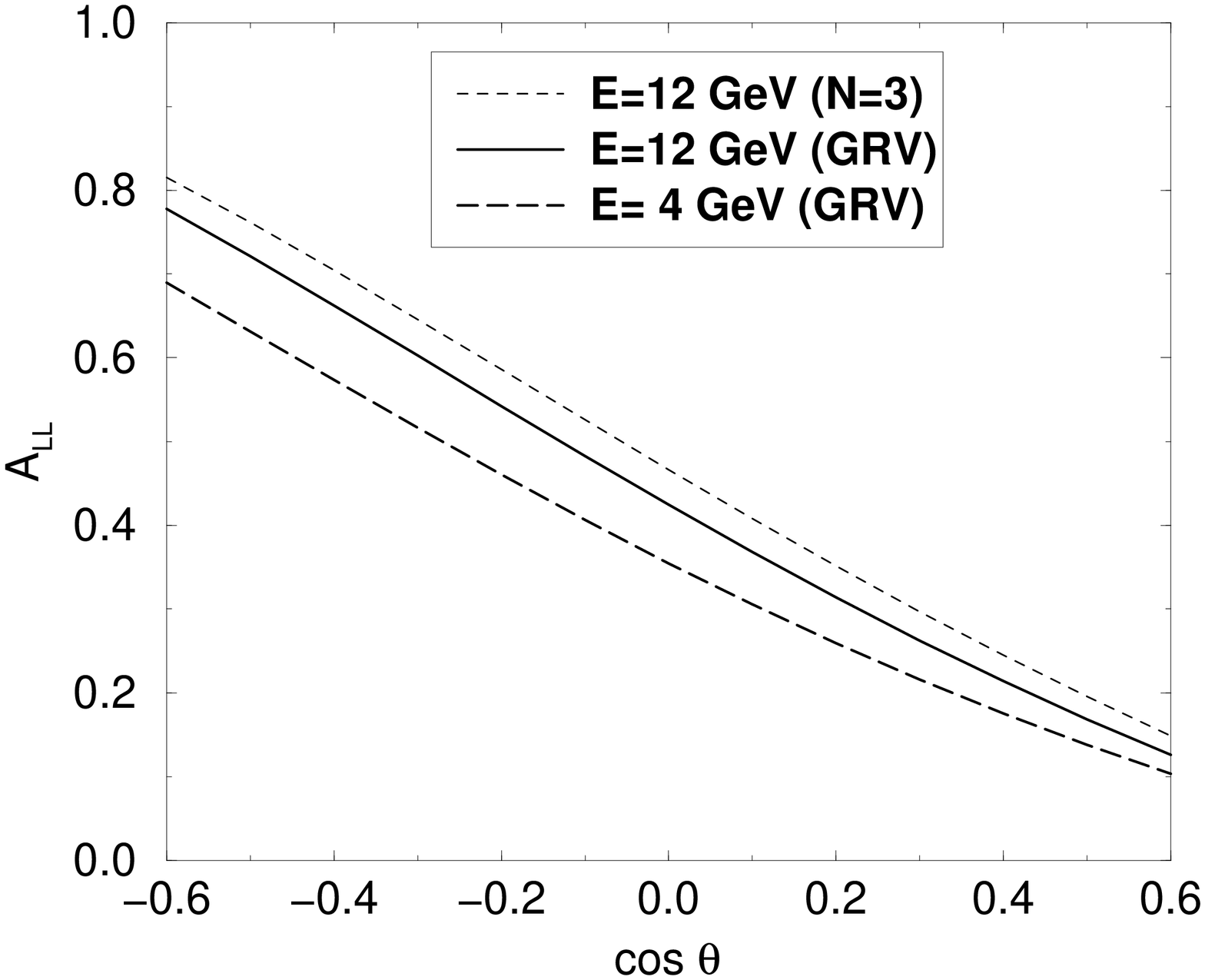, 
          width=5.7cm, angle=-90} 
\vspace*{-0.5cm}
\end{center}}
\caption{The Compton cross section, scaled by $s^{_6}$, (left) 
    and the helicity correlation $A_{\rm LL}$ (right) 
    as predicted by the soft physics approach \cite{DFJK1,DFJK2}. Data taken
    from \cite{shu79}. } 
\vspace*{-0.5cm}
\label{fig:all}
\end{figure}

The amplitude (\ref{final}) leads to the real Compton cross section    
\begin{equation}
\frac{{\d} \sigma}{{\d} t} = \frac{{\d} \hat{\sigma}}{{\d} t}
                       \left [\, \frac{1}{2} (R_V^2(t) + R_A^2(t))
        -\, \frac{us}{s^2+u^2}\, (R_V^2(t)-R_A^2(t)) \,\right] \,.
\label{eq:rcs-cs}
\end{equation}
It is given by the Klein-Nishina cross section, ${\d} \hat{\sigma}/{\d} t$, 
multiplied by a factor that describes the structure of the nucleon in
terms of two form factors. In view of the behaviour of the form factors 
(see Fig.\ \ref{fig:soft}) one infers from (\ref{eq:rcs-cs}) that, in the
soft physics approach, the Compton cross section approximately respects
dimensional counting rule behaviour ($\propto s^{-6}$ at fixed cm
scattering angle $\theta$) in a limited range of energy. The magnitude 
of the Compton cross section is fairly well predicted as is revealed
by comparison with the admittedly old data \cite{shu79} measured at 
rather low values of $s$, $-t$ and $-u$ (see Fig.\ \ref{fig:all}). 
A cross section of similar magnitude has been obtained within the 
diquark model \cite{kro91}, a variant of the standard perturbative 
approach \cite{bro80} in which diquarks are considered as 
quasi-elementary constituents of the proton. It seems difficult for
the perturbative approach to account for the Compton data even if
strongly aymmetric DAs are used \cite{brooks}. Better data are needed
for a crucial test of the soft physics approach and its confrontation
with other approaches. 
The soft physics approach also predicts characteristic spin dependences of the
Compton process. As an example predictions for
the initial state helicity correlation 
\begin{equation}
A_{\rm LL}\, \frac{{\d} \sigma}{{\d} t} = \frac{2\pi\alpha_{\rm em}^2}{s^2} \;
  R_V(t) R_A(t) \left(\frac{u}{s} - \frac{s}{u}\right) \,.
\end{equation}
are shown in Fig.\ \ref{fig:all}.

The cross sections for virtual Compton scattering have been calculated in
\cite{DFJK2}. Characteristic differences to the only other available
results, namely those from the diquark model \cite{kro91}, are to be
noticed. Thus, for instance, the beam asymmetry for $ep\to ep\gamma$
which is sensitive to the imaginary part of the
longitudinal-transverse interference, is zero in the soft physics
approach since all amplitudes are real. In the diquark model, on the
other hand, this asymmetry is non-zero due to perturbatively generated
phases of the Compton amplitudes. In regions
of strong interference between the Compton and the Bethe-Heitler
amplitudes the beam asymmetry is even spectacularly enhanced.

\begin{figure}[hbtp]
\begin{center}
\psfig{file=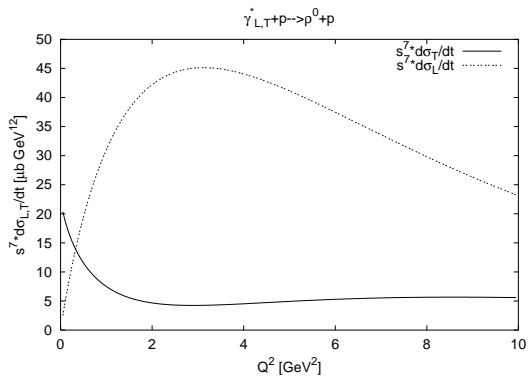,width=5cm,angle=-90}
\end{center}
\caption{The transverse and longitudinal cross sections for the
photoproduction of longitudinally polarised $\rho^0$ mesons at 
$s=40\,\gev^2$ and a cm.\ scattering angle of $90^{\circ}$.}
\label{fig:LT}
\end{figure}

Along the same lines as Compton scattering wide-angle
electroproduction of mesons has also been calculated \cite{hanwen00}.
As an example results for the $\rho^0$-production cross sections for
longitudinally and transversally polarised photons are shown in Fig.\
\ref{fig:LT}. Similarly to deeply virtual electroproduction of mesons
the longitudinal cross section dominates except for $Q^2 \lsim 1
\gev^2$. Contrary to a statement to be found in the literature
occasionally the soft physics approach leads to a $s^7$-scaling of the
cross sections provided $t/s$ and $Q^2/s$ are kept fixed and to the
extent that the form factors $R^M_{V(A)}$ behave $\propto
1/t^2$. Unfortunately there is no wide-angle electroproduction data
available as yet to compare with. For more details and predictions I
refer to \cite{hanwen00}.  

\section{Summary}
The SPDs, generalised PDFs, are new tools for the description of
soft hadronic matrix elements. They are central elements which connect
many different inclusive and exclusive processes: polarised and
unpolarised PDFs are the $\zeta=t=0$ limits of SPDs, electromagnetic
and Compton form factors represent moments of the SPDs, deeply virtual
Compton scattering and hard meson electroproduction are controlled by
them. A particularly interesting aspect is touched in exclusive
reactions such as proton form factors and wide-angle Compton scattering. 
Their analysis by means of SPDs implies the calculation of soft physics
contributions to these processes in which only one of the quarks is
considered  as active while the others act as spectators. The soft
contributions formally represent power corrections to the asymptotically 
leading perturbative contributions in which all quarks participate in 
the subprocess. It seems that for momentum transfers around 10 \gev$^2$ 
the soft contribution dominates over the perturbative one. However, a
severe confrontation of this approach with accurate data on
wide-angle Compton scattering and electroproduction of mesons is pending.

\end{document}